\documentclass[twocolumn,showpacs,preprintnumbers,amsmath,amssymb]{revtex4}
\usepackage{amsmath,amsfonts,amssymb,graphics,graphicx,epsfig,color,times,bbm}

\usepackage{dcolumn}% Align table columns on decimal point
\begin{document}

\preprint{APS/123-QED}

\title{Quantum emitters coupled to surface plasmons of a nano-wire:
A Green function approach
}% Force line breaks with \\

\author{David Dzsotjan$^{1, 2}$}
 %\altaffiliation[Also at ]{}%Lines break automatically or can be forced with \\
\author{Anders S. S{\o}rensen$^{3}$}
\author{Michael Fleischhauer$^{1}$}%

\affiliation{$^{1}$ Department of Physics and research center OPTIMAS, University of Kaiserslautern, Germany;\\ $^{2}$ Research Institute for Particle and Nuclear Physics, H-1525 Budapest, Hungary\\
$^{3}$ QUANTOP, Danish National Research Foundation Center for Quantum Optics, Niels Bohr Institute, DK-2100 Copenhagen, Denmark}%


\date{\today}% It is always \today, today,

\begin{abstract}
We investigate a system consisting of a single, as well as two emitters strongly coupled to surface plasmon modes of a nano-wire using a Green function approach. Explicit expressions are derived for the spontaneous decay rate into the plasmon modes and for the atom-plasmon coupling as well as a plasmon-mediated atom-atom coupling. Phenomena due to the presence of losses in the metal are discussed. In case of two atoms, we observe Dicke sub- and superradiance resulting from their plasmon-mediated interaction. Based on this phenomenon, we propose a scheme for a deterministic two-qubit quantum gate. We also discuss a possible realization of interesting many-body Hamiltonians, such as the spin-boson model, using strong emitter-plasmon coupling.
\end{abstract}

\pacs{Valid PACS appear here}% PACS, the Physics and Astronomy
\maketitle

%%%%%%%%%%%%%%%%%%%%%%%%%%%%%%%%%%%%%%%%%%%%%%%%%%%%%%%%%%%%%%%%%%%%%%%%%%%%%%%

\section{Introduction}

A strong coupling between individual quantum emitters and photons is one of the key ingredients for
photon-based quantum information processing. It is needed to achieve a reliable transfer of excitation between stationary and flying qubits and to realize quantum logic operations. Strong radiative coupling requires a tight confinement of the electromagnetic field. This can be achieved e.g. when quantum emitters are coupled to surface plasmons of nano-wires as proposed in \cite{Chang-PRL-2006}. Besides its simple structure, the atom--nano-wire scheme guarantees an exceptionally strong field-emitter coupling. The original setup involves a single atom placed close to the surface of a metallic wire having sub-wavelength radius. In the present paper we analyze this scheme in detail by means of a Green function approach taking into account metal losses and extend it to a plasmon-mediated coupling between different atoms.  The unique properties of the propagating electromagnetic modes of the wire, the surface plasmons, have opened up new possibilities in many fields such as waveguiding below the diffraction limit \cite{Takahara-OptLett-1998}, sub-wavelength imaging \cite{Klimov-ChemPhysLett-2002} or enhanced fluorescence \cite{Kneipp-PRL-1997}. The small transverse-mode area also enables one to strongly couple an emitter to the plasmon, giving rise to a substantial Purcell effect with a Purcell factor of $10^2$ or larger. The system has been proposed as an efficient single-photon generator \cite{Chang-PRL-2006} as well as a single-photon transistor \cite{Chang-NaturePhys-2007}, strong emitter-plasmon coupling has been experimentally shown using quantum dots \cite{Akimov-Nature-2007}, single plasmons along the wire have been detected, using N-V centers as emitters \cite{Falk-NaturePhys-2009} and the wave-particle duality of surface plasmons has been verified in experiment \cite{Kolesov-NaturePhys-2009}.

We describe a system where a single, as well as two emitters interact with the surface plasmons of the wire in a fully quantum mechanical approach. Due to the strong coupling, the plasmons mediate an intensive long-range interaction between different emitters. As a result, we 
observe sub- and superradiance in a two-atom system which can be efficiently controlled by adjusting the distance between the atoms. Based on this phenomenon, we propose a possible scheme for a deterministic quantum phase gate involving 3-level lambda atoms and classical $2\pi$ laser pulses.

The paper is organized as follows: In Section \ref{sec:green_tensor_guided_modes}  we introduce the Green tensor of the system as well as the structure and properties of the guided modes. Section \ref{sec:coupling_constants} discusses the coupling of a single emitter to the plasmons.  The total spontaneous decay rate and the decay rate into the plasmon modes are calculated. We also derive an expression for the atom-plasmon coupling strength. Section \ref{sec:two_emitters} deals with the long-range interaction of two emitters coupled to the plasmon modes. We show the emergence of Dicke- sub- and superradiance which can be controlled by adjusting the distance between the emitters. We present a scheme for a deterministic quantum phase gate based on the controllable superradiance phenomenon. Both in the case of a single emitter, as well as a pair of them, we discuss the effects of metal losses and their consequences. Finally, we discuss a possible extension of the level scheme of the emitters to realize interesting many-body Hamiltonians such as spin-boson model using the coupling to nanowires. Section \ref{sec:summary} contains a summary of the results.

%%%%%%%%%%%%%%%%%%%%%%%%%%%%%%%%%%%%%%%%%%%%%%%%%%%%%%%%%%%%%%%%%%%%%%%%%%%%%%%%%%%%%
%%%%%%%%%%%%%%%%%%%%%%%%%%%%%%%%%%%%%%%%%%%%%%%%%%%%%%%%%%%%%%%%%%%%%%%%%%%%%%%%%%%%%
\section{Electromagnetic Green tensor and guided modes of the wire}\label{sec:green_tensor_guided_modes}
%%%%%%%%%%%%%%%%%%%%%%%%%%%%%%%%%%%%%%%%%%%%%%%%%%%%%%%%%%%%%%%%%%%%%%%%%%%%%%%%%%%%%
%%%%%%%%%%%%%%%%%%%%%%%%%%%%%%%%%%%%%%%%%%%%%%%%%%%%%%%%%%%%%%%%%%%%%%%%%%%%%%%%%%%%%

The coupling of quantum emitters to the quantized electromagentic field in the
vicinity of linear macroscopic media can be described by incorporating the effect of the
medium in the Green tensor of the field
$\bar{\bar{G}}(\vec{r},\vec{r}^{\, \prime},\omega)$  which
obeys the Maxwell-Helmholtz wave equation with a Dirac delta source at ${\vec r}=\vec{r}^{\, \prime}$
\begin{equation}
 \left[\nabla\times\nabla\times-\frac{\omega^2}{c^2}\epsilon(\vec{r},\omega)\right]\bar{\bar{G}}(\vec{r},
\vec{r}^{\, \prime},\omega)=\bar{\bar{I}}\,\delta(\vec{r}-\vec{r}^{\, \prime}).
\end{equation}
Here $\epsilon({\vec r},\omega)$ is the
complex dielectric function of the isotropic linear medium at position $\vec{r}$ and frequency $\omega$. In the following, we will use $\epsilon=-75+0.6i$ as typical value which corresponds to the electric permittivity of silver at around $\lambda=1\mu m$.
It is assumed that the magnetic response is negligible. 

In the following we will calculate the Green tensor for a nano-wire.
We consider a cylindrical wire with radius $R$ and symmetry axis $z$. The
metal is described by a complex dielectric function with $\epsilon^\prime=\mathrm{Re}[\epsilon]<0$. For simplicity
we assume an infinitly long wire to have translational invariance along the $z$-axis. For
$\vec{r}^{\, \prime}$ being outside the cylinder we can decompose the Green tensor in the following way:
\begin{equation}
\bar{\bar{G}}(\vec{r},\vec{r}^{\, \prime},\omega)= 
\begin{cases}
 {\bar{\bar{G}}_0(\vec{r},\vec{r}^{\, \prime},\omega)+\bar{\bar{G}}_R(\vec{r},\vec{r}^{\, \prime},\omega)} & {r>R }\\
{\bar{\bar{G}}_T(\vec{r},\vec{r}^{\, \prime},\omega)} & {r<R }
\end{cases}
\end{equation}
where, if the observation point $\vec{r}$ is also outside the cylinder, we have the sum of a direct (or vacuum) and a reflected contribution, whereas inside the cylinder we get only a transmitted term. $r$ is the distance from the wire axis.
To fulfill the boundary conditions, we expand the Green tensor in cylindrical harmonics \cite{Li-JElMagnWavesAppl-2000}. The cylindrical harmonic vector wave functions are given by
\begin{eqnarray}
\left\{ ^{\vec{M}_{e,n}(k_z,\vec{r})}_{\vec{M}_{o,n}(k_z,\vec{r})}\right\}&=&\left\{^{\nabla\times\left[Z_n(k_{r_{0,1}}r)\left(\cos n\phi\right)e^{ik_zz}\hat{z}\right]}_{\nabla\times\left[Z_n(k_{r_{0,1}}r)\left(\sin n\phi\right) 
e^{ik_zz}\hat{z}\right]}\right\}\nonumber\\
\left\{
^{\vec{N}_{e,n}(k_z,\vec{r})}_{\vec{N}_{o,n}(k_z,\vec{r})}\right\}&=&\left\{^{\frac{1}{k_{0,1}}\nabla\times\vec{M}_{e,n}(k_z,\vec{r})}_{\frac{1}{k_{0,1}}\nabla\times\vec{M}_{o,n}(k_z,\vec{r})}\right\},
\end{eqnarray}
with $k_0=\frac{\omega}{c}$, $k_1=\sqrt{\epsilon_1}k_0$ and $k_{r_{0,1}}^2=k_{0,1}^2-k_z^2$ where we have $0,1$ for outside and inside the cylinder and the subscripts $e$ and $o$ stand for even and odd, respectively. Note that $k_1$ is complex due to
the non-vanishing imaginary part of the electric 
%04022010 corrected \text (came out wrong when I compiled)
permittivity $\epsilon^{\prime\prime}=\text{Im}[\epsilon]$.
The radial part $Z_n(x)$ has to be either replaced by the Bessel function $J_n(x)$, or, 
%04022010 Tried to clarify a bit here
when the ``$(1)$'' appears as a superscript on $\vec{M}$ or $\vec{N}$, by $H_n^{(1)}(x)$, i.e. the Hankel function of the first kind. 
Note, furthermore, that the tensorial products between the even and odd cylindrical vector wave functions are defined according to $\vec{M}_{^e_on}\vec{N}^{(1)}_{^o_en}=\vec{M}_{e,n}\vec{N}^{(1)}_{o,n}+\vec{M}_{o,n}\vec{N}^{(1)}_{e,n}$ and all other combinations similarly. For better readability, we have omitted the tensorial product symbol 
between them.

This gives for the direct term of the Green tensor
\begin{eqnarray}
&&\bar{\bar{G}}_0(\vec{r},\vec{r}^{\, \prime},\omega)= \nonumber\\
&& -\frac{\hat{r}\hat{r}\delta(\vec{r}-\vec{r}^{\,\prime})}{k_0^2}
+\frac{i}{8\pi}\int_{-\infty}^{\infty}\!\!\!\!dk_z\sum_{n=0}^{\infty}\frac{2-\delta_{n,0}}{k_{r_0}^2} \\
&&\times
 \begin{cases} 
\vec{M}^{(1)}_{^e_o n}(k_z,\vec{r})\vec{M}_{^e_o n}(-k_z,\vec{r}^{\, \prime})+\vec{N}^{(1)}_{^e_o n}(k_z,\vec{r})\vec{N}_{^e_o n}(-k_z,\vec{r}^{\, \prime})\\ 
\vec{M}_{^e_o n}(k_z,\vec{r})\vec{M}^{(1)}_{^e_o n}(-k_z,\vec{r}^{\, \prime})+\vec{N}_{^e_o n}(k_z,\vec{r})\vec{N}^{(1)}_{^e_o n}(-k_z,\vec{r}^{\, \prime}) 
\end{cases}\nonumber \label{Eq:Green_vac}
\end{eqnarray}
where the upper (lower) line holds for $r>r^\prime$ ($r <r^\prime$).
For the reflected part we find
\begin{eqnarray}
\begin{aligned}
&\bar{\bar{G}}_R(\vec{r},\vec{r}^{\, \prime},\omega)= \\
& \frac{i}{8\pi}\int_{-\infty}^{\infty}\!\!\!\!dk_z\sum_{n=0}^{\infty}\frac{2-\delta_{n,0}}{k_{r_0}^2} \\
& [\left(A_R\vec{M}^{(1)}_{^e_on}(k_z,\vec{r})+B_R\vec{N}^{(1)}_{^o_en}(k_z,\vec{r})\right)\vec{M}^{(1)}_{^e_on}(-k_z,\vec{r}^{\, \prime}) \\
& +\left(C_R\vec{N}^{(1)}_{^e_on}(k_z,\vec{r})+D_R\vec{M}^{(1)}_{^o_en}(k_z,\vec{r})\right)\vec{N}^{(1)}_{^e_on}(-k_z,\vec{r}^{\, \prime})].
\end{aligned}\label{Eq:Green_refl}
\end{eqnarray}
Finally, we have for the transmitted part
\begin{eqnarray}
\begin{aligned}
&\bar{\bar{G}}_T(\vec{r},\vec{r}^{\, \prime},\omega)= \\
& \frac{i}{8\pi}\int_{-\infty}^{\infty}\!\!\!\!dk_z\sum_{n=0}^{\infty}\frac{2-\delta_{n,0}}{k_{r_0}^2} \\
& [\left(A_T\vec{M}_{^e_on}(k_z,\vec{r})+B_T\vec{N}_{^o_en}(k_z,\vec{r})\right)\vec{M}^{(1)}_{^e_on}(-k_z,\vec{r}^{\, \prime}) \\
& +\left(C_T\vec{N}_{^e_on}(k_z,\vec{r})+D_T\vec{M}_{^o_en}(k_z,\vec{r})\right)\vec{N}^{(1)}_{^e_on}(-k_z,\vec{r}^{\, \prime})].
\end{aligned}\label{Eq:Green_trans}
\end{eqnarray}
The reflection and transmission coefficients  $A_{R,T}, B_{R,T}, C_{R,T}, D_{R,T}$ can be specified by imposing the boundary conditions at the surface of the cylinder ($r=R $):
\begin{eqnarray}
 \hat{r}\times\bar{\bar{G}}(\vec{r},\vec{r}^{\, \prime})_{r=R ^-}&=&\hat{r}\times\bar{\bar{G}}(\vec{r},\vec{r}^{\, 
\prime})_{r=R ^+} \label{eq:boundary1}\\
\hat{r}\times\nabla\times\bar{\bar{G}}(\vec{r},\vec{r}^{\, \prime})_{r=R ^-}&=&\hat{r}\times\nabla\times\bar{\bar{G}}(\vec{r},\vec{r}^{\, \prime})_{r=R ^+}.\label{eq:boundary2}
\end{eqnarray}
Solving the above equations, we get the full Green tensor. 

%%%%%%%%%%%%%%%%%%%%%%%%%%%%%%%%%%%%%%%%%%%%%%%%
\subsection{longitudinal modes}
%%%%%%%%%%%%%%%%%%%%%%%%%%%%%%%%%%%%%%%%%%%%%%%%%

\begin{figure}
 \includegraphics[width=9cm,angle=0]{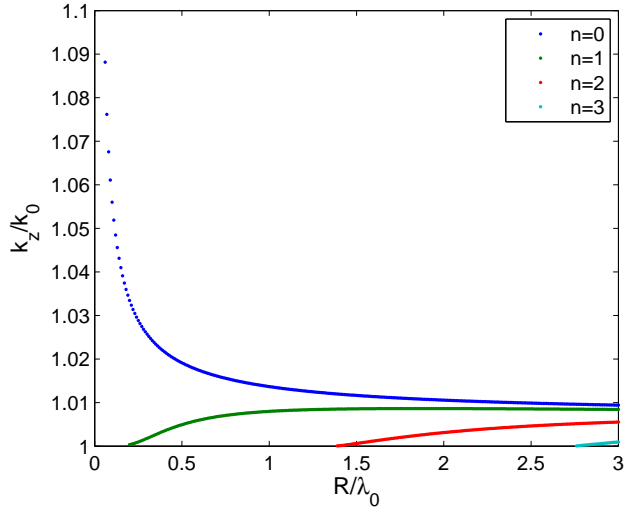}
\caption{\label{fig:allowedmodes} $k_z$ components of the guided modes of a nanowire ($\epsilon=-75$), as the function of the wire radius $R$ and cylindrical harmonic order $n$. %04022010 rewrote a bit
When the radius  decreases, all modes with a non-zero $n$ have a cutoff, whereas the $n=0$ mode has an increasing $k_z$, that is, an icreasingly strong confinement. $R$ is  scaled with the vacuum radiation wavelength $\lambda_0$.}
\end{figure}

Let us first discuss the solution of the boundary equations in the absence of
material losses, i.e. for $\epsilon^{\prime\prime}=0$. Writing up (\ref{eq:boundary1}) and (\ref{eq:boundary2}), one finds a system of linear algebraic equations for the transmission and reflection coefficients $A_{R,T}, B_{R,T}, C_{R,T}, D_{R,T}$. One can rewrite these as a vector, containing the coefficients, multiplied by a matrix.
In order for the equations to have a nontrivial solution infinitely far from the source (i.e where we can neglect the contribution from $\bar{\bar{G_0}}$), the determinant of the matrix has to vanish. From this condition we obtain a mode equation for each harmonic order $n$. These mode
equations are equivalent to those obtained by different methods in \cite{Chang-PRL-2006}:
\begin{eqnarray}\label{eq:mode_eq}
\begin{aligned}
 &\frac{n^2k_z^2}{R ^2}\left(\frac{1}{k_{r_1}^2}-\frac{1}{k_{r_0}^2}\right)^2=\\
&\quad\left(\frac{1}{k_{r_1}}\frac{J_n^{\, \prime}(k_{r_1}R )}{J_n(k_{r_1}R )}-\frac{1}{k_{r_0}}\frac{H_n^{\, \prime}(k_{r_0}R )}{H_n(k_{r_0}R )}\right)\\
&\times\left(\frac{k_1^2}{k_{r_1}}\frac{J_n^{\, \prime}(k_{r_1}R )}{J_n(k_{r_1}R )}-\frac{k_0^2}{k_{r_0}}\frac{H_n^{\, \prime}(k_{r_0}R )}{H_n(k_{r_0}R )}\right)
\end{aligned}
\end{eqnarray}
Fixing the frequency $\omega$, we get an equation for the longitudinal component of the wavevector parallel to the wire, the roots giving the allowed values $k_z=k_z(\omega;n)$ for each transversal mode $n$. These are the surface plasmons. In Fig.\ref{fig:allowedmodes} the allowed values of $k_z$ are plotted as function of wire radius $R$ for the different cylindrical harmonics.  One recognizes
that with the shrinking of $R$, all solutions belonging to $n\ne0$ modes have a cutoff and only the one belonging to $n=0$ persists. The corresponding value of $k_z$ diverges as $R\rightarrow 0$ .

When calculating the reflected and transmitted Green tensors, as seen in (\ref{Eq:Green_refl}) and (\ref{Eq:Green_trans}), the terms inside the integral and the sum on their right hand sides have a singularity exactly at $k_z=k_z(\omega;n)$. This means that there is a contribution to the response of the nano-wire system that originates from the $n$-th plasmonic mode.

%%%%%%%%%%%%%%%%%%%%%%%%%%%%%%%%%%%%%%%%%%%%%%%%%%%%%%%%%%%%%%%%%%%%%%%%%%%%%%%%%%%%%%%%%%%%%%%%%%%%%%%%%%%%%%%%%%%%%%%
%%%%%%%%%%%%%%%%%%%%%%%%%%%%%%%%%%%%%%%%%%%%%%%%%%%%%%%%%%%%%%%%%%%%%%%%%%%%%%%%%%%%%%%%%%%%%%%%%%%%%%%%%%%%%%%%%%%%%%%%

%%%%%%%%%%%%%%%%%%%%%%%%%%%%%%%%%%%%%%%%%%%%%%%%%%%%%%%%%%%
\subsection{width of the plasmon resonances}\label{subsec:resonance_width}
%%%%%%%%%%%%%%%%%%%%%%%%%%%%%%%%%%%%%%%%%%%%%%%%%%%%%%%%%%%

Taking into account the medium absorption, i.e. a small but finite value of $\epsilon^{\prime\prime}$, the discrete cylindrical modes turn into resonances. This means that in case of losses there is no longer a single, well-defined $k_z$ but continuously many $k_z$ values peaked around $k_z(\omega;n)$ that contribute to the plasmon mode. The consequences can be drawn much clearer if the wire radius is much smaller than the optical wavelength, because then we can interpret the propagation of the plasmon mode as a 1D problem, looking at a single mode only. 
The material losses induce a distribution of components with different $k_z$ values in the propagating plasmonic mode, each acquiring a phase factor $exp(ik_zz)$ while propagating a distance $z$. This, in turn, results in a dephasing of the components upon propagation, %04022010  added a little here
which describes the losses.

\begin{figure}
 \includegraphics[width=9cm, angle=0]{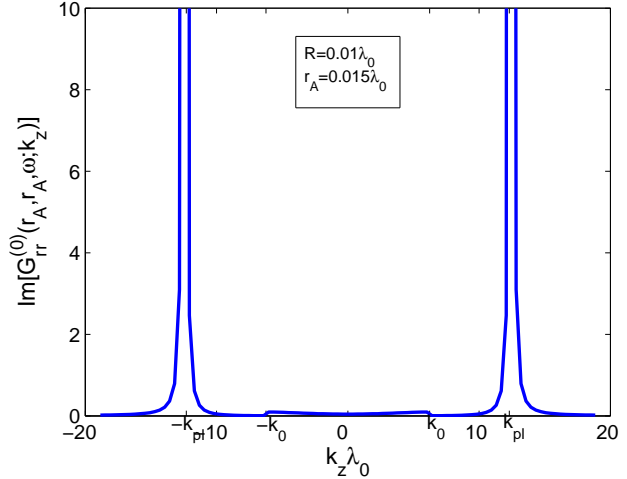}
\caption{\label{fig:plasmon_resonance} $\mathrm{Im}\left[\hat{r}^T\bar{\bar{G}}^{(n=0)}(\vec{r}_A,\vec{r}_A,\omega;k_z)\hat{r}\right]$, in case of a single-mode, lossy wire ($\epsilon=-75+0.6i$). $\hat{r}$ is the unit vector in the radial direction. The broadened plasmonic resonance peaks show up at around the plasmonic longitudinal wavenumbers $\pm k_{pl}$ (predicted by (\ref{eq:mode_eq})), in the evanescent region ($k_{pl}>k_0$). The distances and wave numbers are scaled with the vacuum radiation wavelength $\lambda_0$.}
\end{figure}

\begin{figure}
 \includegraphics[width=9cm,angle=0]{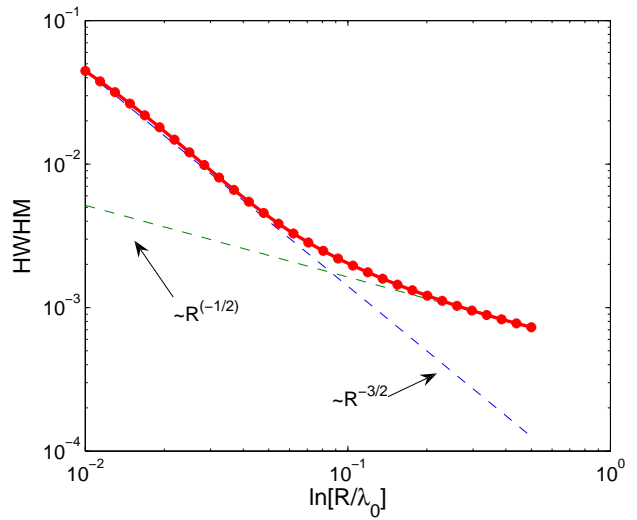}
\caption{\label{fig:resonance_width} HWHM of the plasmonic resonance peaks shown in Fig.\ref{fig:plasmon_resonance}, as a function of wire radius $R$, in a log-log plot. The width and the radius are scaled by vacuum radiation wavelength $\lambda_0$. The electric permittivity is $\epsilon=-75+0.6i$. As the wire gets thinner, the broadening of the resonance increases, %04022010 changed here - I find it counter intuitive to there is loss because it get broader. I would say it is the other way around
i.e., the propagation losses increases.
% which results in increasing propagation losses for the plasmonic modes.
}
\end{figure}

Fig. \ref{fig:plasmon_resonance} shows the shape of the resonance belonging to the $n=0$ cylindrical mode around $k_z=\pm k_z(\omega;n)$, the value corresponding to the lossless case. The lineshape is well approximated with a Lorentzian function. The HWHM of the resonance peak scales linearly with $\epsilon^{\prime\prime}$, as one would expect. In Fig.\ref{fig:resonance_width} the value of the HWHM is depicted, which assumes larger and larger values as the wire radius $R$ gets smaller. One recognizes a transition from a slower $R^{-1/2}$ to a faster $R^{-3/2}$ dependence of the  width of the plasmon resonances at around $R\sim10^{-1}\lambda_0$.

So, in case of $\epsilon^{\prime\prime}\ne0$, decreasing the wire radius is accompanied not only by a decreasing transverse mode area, but also by increasing propagation losses. The two effects are connected because the smaller the transverse mode area gets, the larger part of the propagating field will be concentrated inside the metal which induces growing propagation losses. We will see later that for a decreasing value of $R$ the detrimental effects of propagation losses will overcome the positive effect of an increased coupling, resulting in an optimal wire radius. 

%%%%%%%%%%%%%%%%%%%%%%%%%%%%%%%%%%%%%%%%%%%%%%%%%%%%%%%%%%%%%
%%%%%%%%%%%%%%%%%%%%%%%%%%%%%%%%%%%%%%%%%%%%%%%%%%%%%%%%%%%%%%
%%%%%%%%%%%%%%%%%%%%%%%%%%%%%%%%%%%%%%%%%%%%%%%%%%%%%%%%%%%%%%

%%%%%%%%%%%%%%%%%%%%%%%%%%%%%%%%%%%%%%%%%%%%%%%%%%%%%%%%%%%%%%%%%%%%%%%%%%%%%%%%%%%%%%%%%%%%%%%%%%%%%
\section{\label{sec:coupling_constants} Atom-plasmon coupling}
%%%%%%%%%%%%%%%%%%%%%%%%%%%%%%%%%%%%%%%%%%%%%%%%%%%%%%%%%%%%%%%%%%%%%%%%%%%%%%%%%%%%%%%%%%%%%%%%%%%%%

\subsection{general framework}

A two-level atom with an electric dipole transition of strength $\vec d$ couples to the guided plasmon waves
via the quantized electromagnetic field $\hat {\vec E}(\vec r_A)$ at the position of the
atom. The interaction Hamiltonian in %04022010 inserted "the" and approximationS
the dipole and rotating wave-approximations in the Schr\"{o}dinger picture reads
\begin{equation}
 \hat H_I= -  \hat \sigma^\dagger \vec d\cdot{ \hat{\vec E}}^{(+)}\!\!(\vec r_A)\, + 
h.a.\label{Eq:dipole_interaction}
\end{equation}
where 
\begin{equation}
 \hat{\vec{E}}^{(+)}(\vec{r})=\int_{0}^{\infty}\!\!{\rm d}\omega\hat{\vec{E}}(\vec{r},\omega).\label{Eq:field_op}
\end{equation}
is the positive frequency part of the electric field operator in the presence of the nano-wire
and $\hat \sigma^\dagger=|e\rangle\langle g|$ is the 
atomic raising operator between the ground state $|g\rangle$ and the excited state $|e\rangle$ of the atom. The negative frequency counterpart of $\hat{\vec{E}}^{(+)}(\vec{r})$ is defined as

\begin {equation}
 \hat{\vec{E}}^{(-)}(\vec{r})=\int_{0}^{\infty}\!\!{\rm d}\omega\hat{\vec{E}}^\dagger(\vec{r},\omega).\label{Eq:field_op_neg}
\end {equation}

Following the treatment of \cite{Dung-PRA-2003}, the electric field operator
\begin{equation}\label{eq:E}
 \hat{\vec{E}}(\vec{r},\omega)=i\sqrt{\frac{\hbar}{\pi\epsilon_0}}\frac{\omega^2}{c^2}\int 
{\rm d}^3\vec{r}^{\, \prime}\sqrt{\epsilon^{\prime\prime}(\vec{r}^{\, \prime},\omega)}\bar{\bar{G}}(\vec{r},\vec{r}^{\, \prime},\omega)\hat{\vec{f}}_{\omega}(\vec{r}^{\, \prime})
\end{equation}
can be expressed in terms of elementary excitations $\hat {\vec f}_{\omega}(\vec r)$ and $\hat {\vec f}_{\omega}^\dagger(\vec r)$ fulfilling bosonic commutation relations
\begin{eqnarray}
 \left[\hat{f}_{\omega^{}_{i}}(\vec{r}),\hat{f}^{\dagger}_{\omega^{\prime}_{j}}({\vec{r}}^{\, \prime})\right]&=&\delta_{ij}
\delta(\vec{r}-\vec{r}^{\, \prime})\delta(\omega-\omega^{\prime}), \\
 \left[\hat{f}_{\omega^{}_i}(\vec{r},\omega),\hat{f}_{\omega^{\prime}_j}({\vec{r}}^{\, \prime},\omega^{\prime})\right]&=&0.
\end{eqnarray}
Here, $\epsilon^{\prime\prime}$ is the imaginary part of the electric permittivity. The interactionless Hamiltonians of the combined field-matter system and a single, two-level atom have the simple form
\begin{eqnarray}\label{eq:interactionless_hamiltonians}
 \hat{H}^F_0&=&\int \!\!{\rm d}^3\vec{r}\int_{0}^{\infty} 
\!\!\! {\rm d}\omega\, \hbar\omega\, \, \hat{\vec{f^{\dagger}_{\omega}}}(\vec{r})\hat{\vec{f}}_{\omega}(\vec{r},\omega),\nonumber\\
\hat{H}^A_0&=&\frac{1}{2}\hbar\omega_A\hat{\sigma}_z
\end{eqnarray}
where $\omega_A$ is the atomic transition frequency and $\hat{\sigma}_z=|e\rangle\langle e|-|g\rangle\langle g|$.

%%%%%%%%%%%%%%%%%%%%%%%%%%%%%%%%%%%%%%%%%%%%%%%%%%%%%%%%%%%%%%%%%%%%%%%%%%%%%%%%%%%%%%%%%%%%%%%%%%%%%%
\subsection{\label{sec:single_atom}  spontaneous emission and level shift of an atom near a nanowire}
%%%%%%%%%%%%%%%%%%%%%%%%%%%%%%%%%%%%%%%%%%%%%%%%%%%%%%%%%%%%%%%%%%%%%%%%%%%%%%%%%%%%%%%%%%%%%%%%%%%%%%

It is well known that a dielectric body near an atom influences its spontaneous emisson rate. The interaction strength of an atom with the different modes of the electromagnetic field is inversely proportional to the square root of the effective mode volume.  That also means that the probability of spontaneous emission into modes with the smallest effective mode volume will be the largest. Additionally, slow group velocity of the given mode also enhances the interaction strength.

\begin{figure}
 \includegraphics[width=9cm,angle=0]{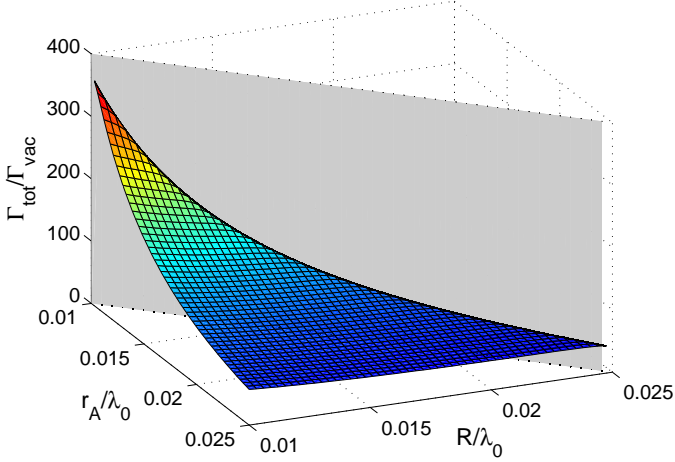}
\caption{\label{fig:spont_em} Spontaneous decay rate of a single emitter relative to the vacuum spontaneous decay rate as a function of emitter - wire axis distance $r_A$ and wire radius $R$, both scaled by vacuum radiation wavelength $\lambda_0$. We use $\epsilon=-75+0.6i$. Because of the strong interaction of the emitter with the wire eigenmodes, there is a considerable increase in the spontaneous decay rate at small radii and emitter-wire distances. Since $\Gamma_{tot}\rightarrow\infty$ for $r_A\rightarrow R$,%04022010 changed here
we have introduced a cutoff at a small $r_A-R$ for better visibility. }
\end{figure}

 In the following we will take a look at the spontaneous emission rate of a two-level atom near the surface of a nanowire. 
We have seen before that it is possible to have a wire with a single, strongly confined mode at a given frequency, that is, a mode with a very small effective cross section area. Thus, we expect that if we place the atom close enough to the wire, its spontaneous emission rate will increase considerably \cite{Chang-PRL-2006}. To ascertain this, we write up the Heisenberg equations of motion for the field and atomic operators. 
\begin{eqnarray}
 \dot{\hat{\sigma}}_z&=&\frac{2i}{\hbar}\hat{\sigma}^\dagger\hat{\vec{E}}^{(+)}(\vec{r}_{A},t)\vec{d}+H.c. \label{eq:sigmamotion1}\\
\dot{\hat{\sigma}}^\dagger&=&i\omega_A\hat{\sigma}^\dagger+\frac{i}{\hbar}\hat{\vec{E}}^{(-)}(\vec{r}_A,t)\vec{d}\hat{\sigma}_z \label{eq:sigmamotion2} 
\end{eqnarray}
where the operator of the electric field is given by (\ref{eq:E}).  The equation of motion of the electric field, respectively the elementary excitations $\hat f$ reads
\begin{eqnarray}
\dot{\hat{\vec{f}}}_{\omega}(\vec{r})&=&-i\omega\hat{\vec{f}}_{\omega}(\vec{r})+\nonumber \\
&&+\frac{\omega^2}{c^2}\sqrt{\frac{\epsilon^{\prime\prime}(\vec{r},\omega)}{\hbar\pi\epsilon_0}}\bar{\bar{G}}^*(\vec{r},\vec{r}_A,\omega)\vec{d}\hat{\sigma} \label{eq:fmotion}
\end{eqnarray}
To eliminate the field, we formally integrate (\ref{eq:fmotion}) and substitute it into (\ref{eq:sigmamotion1}) and (\ref{eq:sigmamotion2}). Employing a Markov approximation we get rid of the time integral, and using the Green tensor property (\cite{Dung-PRA-2003})
\begin{eqnarray}
&&\int d^3r^{\prime} \frac{\omega^2}{c^2}\epsilon^{\prime\prime}(\vec{r}^{\prime},\omega)\bar{\bar{G}}(\vec{r}_1,\vec{r}^{\prime},\omega)\bar{\bar{G}}^{\dagger}(\vec{r}_2,\vec{r}^{\prime},\omega)\nonumber \\
&&=\mathrm{Im}\left[\bar{\bar{G}}(\vec{r}_1,\vec{r}_2,\omega)\right],\label{Eq:Green_property}
\end{eqnarray}
we obtain the effective equations for $\hat{\sigma}_z$ and $\hat{\sigma}^\dagger$:
\begin{eqnarray}
 \dot{\hat{\sigma}}_z&=&-\Gamma_{tot}(1+\hat{\sigma}_z)+\left[\frac{2i}{\hbar}\hat{\sigma}^\dagger\hat{\vec{E}}^{(+)}_{free}\vec{d}+H.c.\right]\\
\dot{\hat{\sigma}}^\dagger&=&\left[i(\omega_A-\delta\omega)-\frac{1}{2}\Gamma_{tot}\right]\hat{\sigma}^\dagger+\frac{i}{\hbar}\hat{\vec{E}}^{(-)}_{free}\vec{d}\hat{\sigma}_z
\end{eqnarray}
where $\hat{\vec{E}}_{free}^{(\pm)}=\hat{\vec{E}}_{free}^{(\pm)}(\vec{r}_{A},t)$ is the free part of the field proportional to the free part of $\hat f$ and $\hat{f}^{\dagger}$.
The total decay rate and the Lamb shift $\Gamma_{tot}$ and $\delta\omega$
can then be expressed in terms of the electromagnetic Green tensor
\begin{eqnarray}\label{eq:decay_lambshift}
 \Gamma_{tot}&=&\frac{2\omega_A^2d_id_j}{\hbar\epsilon_0c^2}\mathrm{Im}\left[G_{ij}(\vec{r}_A,\vec{r}_A,\omega_A)\right]\\
\delta\omega&=&\frac{d_id_j}{\hbar\epsilon_0\pi}\mathcal{P}\int_{0}^{\infty}\!\!\!\!d\omega\frac{\omega^2}{c^2}\frac{\mathrm{Im}\left[G_{ij}(\vec{r}_A,\vec{r}_A,\omega)\right]}{\omega-\omega_A}.
\end{eqnarray}
Fig.~\ref{fig:spont_em} shows the results of the numerical simulation for $\Gamma_{tot}$ for different atom - wire axis distances and wire radii. According to these results , the interaction of the atom with the surface plasmon mode gives rise to a significant Purcell effect, enhancing the spontaneous emission rate by two orders of magnitude. Thus, the atom interacts with the surface plasmon mode in the strong coupling regime. It is important to note that the interaction of the atom with the wire strongly depends on the direction of the polarization vector $\vec{d}$. The coupling to the plasmons is strongest when $\vec{d}$ points in the radial direction. In the simulation we used this configuration.
%%%%%%%%%%%%%%%%%%%%%%%%%%%%%%%%%%%%%%%%%%%%%%%%%%%%%%%%%%%%%%%
%%%%%%%%%%%%%%%%%%%%%%%%%%%%%%%%%%%%%%%%%%%%%%%%%%%%%%%%%%%%%%%%

\begin{figure}
 \includegraphics[width=9cm,angle=0]{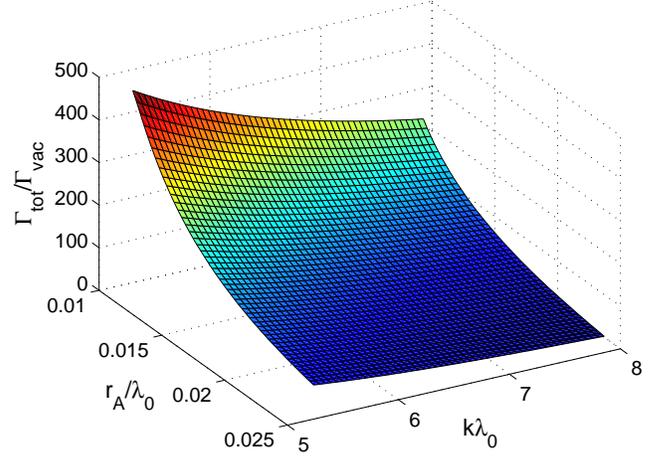}
\caption{\label{fig:spont_em_spectrum} Spontaneous decay rate of a single emitter relative to the vacuum spontaneous decay rate as a function of wave number $k=\omega/c$ and emitter - wire axis distance $r_A$, in case of a lossy wire ($\epsilon=-75+0.6i$) with radius $R=0.01\lambda_0$. Both $k$ and $r_A$ are scaled by vacuum radiation wavelength $\lambda_0$. The large width in frequency verifies the applicability of %04022010 inserted 
 the Markov approximation. Since $\Gamma_{tot}\rightarrow\infty$ for $r_A\rightarrow R$, %04022010 inserted "have"
%when plotting $\Gamma_{tot}/\Gamma_{vac}$ 
we have introduced a cutoff at a small $r_A-R$ for better visibility.}
\end{figure}

%Anders edited here: it is not really a spectrum so I don't like calling it that
We have yet to justify the Markov approximation. In Fig.~\ref{fig:spont_em_spectrum} we show  $\Gamma_{tot}$, as a function of different atom-wire distances and frequency, taking a linear dispersion (this is a good approximation at optical frequencies). The results show that  $\Gamma_{tot}$ has a very large width in frequency, much more than $0.5 \omega_A$, which means that the recurrence time (which is proportional to the inverse of the width) is several orders of magnitude smaller than the characteristic time scale of the atomic decay. %04022010 inserted: 
This justifies the use of the Markov approximation

%%%%%%%%%%%%%%%%%%%%%%%%%%%%%%%%%%%%%%%%%%%%%%%%%%%%%%%%%%%%%%%%%%%
%%%%%%%%%%%%%%%%%%%%%%%%%%%%%%%%%%%%%%%%%%%%%%%%%%%%%%%%%%%%%%%%%%%%

%%%%%%%%%%%%%%%%%%%%%%%%%%%%%%%%%%%%%%%%%%%%%%%%
\subsection{sub-wavelength nano-wire}\label{subsec:sub-wavelength wire}
%%%%%%%%%%%%%%%%%%%%%%%%%%%%%%%%%%%%%%%%%%%%%%%%%

Although with the Green tensor formalism one can handle the problem in an elegant way, it describes the interaction of one or many emitters with their surroundings in a compact form where contributions of different effects are difficult to distinguish. For example, when calculating the spontaneous emission of a single atom, we obtain the total emission rate into all possible modes of the environment. Thus, on first glance, the result  does not enable us to separately determine the rate of spontaneous emission into electromagnetic modes associated with the nano-wire plasmons and those into the radiative modes. In the following we will show, however, that with some further calculations we can still resolve the contribution of the different modes. 
We can define the following Hamiltonian:
\begin{equation}
\hat{H}_I=\hat{H}^{tr}_I+\hat{H}^{ev}_I
\end{equation}
%
%
%04022010 change here
where we have separated the interaction into the evanescent modes $(|k_z|>k_0)$ and the traveling-wave part $(|k_z|\leq k_0)$ of the electromagnetic field. We do this because, as one can see in Fig.\ref{fig:allowedmodes}, the plasmons are evanescent in the radial direction $(|k_z|>k_0)$, never appearing in the traveling-wave part of the $k$-spectrum. Moreover, %04022010 
for lossless media $\epsilon^{\prime\prime}=0$, one can verify from the structure of $\mathrm{Im}[\bar{\bar{G}}^{(n)}(\vec{r}_A,\vec{r}_A,\omega;k_z)]$ that the surface plasmons are the only contribution in the evanescent region. In this way we can define traveling and evanescent creation and annihilation operators $\hat{a}_\omega^{\dagger tr,ev},\hat{a}_\omega^{tr,ev}$ with the usual commutation relations, so that
\begin{equation}
 \hat{H}_I=-\int_0^{\infty}\!\!\!\!d\omega\, g^{tr}(\omega)\hat{a}_\omega^{tr}\hat{\sigma}^\dagger-\int_0^{\infty}\!\!\!\!d\omega\, g^{ev}(\omega)\hat{a}_\omega^{ev}\hat{\sigma}^\dagger+h.a.
\end{equation}
where $g^{ev,tr}(\omega)$ is the coupling strength to the evanescent and traveling modes, respectively, at frequency $\omega$. $\hat{\sigma}^\dagger$ is the atomic flip operator. 
Writing up the Heisenberg equation for $\hat{a}^{ev,tr}_{\omega}$ and $\hat{\sigma}$, and using that the traveling and evanescent operators commute, we get
\begin{eqnarray}
 \dot{\hat{a}}^{ev,tr}_{\omega}&=&-i\omega\hat{a}^{ev,tr}_{\omega}(t)+\frac{i}{\hbar}g^{ev,tr}(\omega)^*\hat{\sigma}(t)\label{eq:eqmotion_a}\\
 \dot{\hat{\sigma}}&=&-i\omega_A\hat{\sigma}(t)\nonumber\\
&&-\frac{i}{\hbar}\int^\infty_{0}\!\!\!\!d\omega\left(g^{tr}(\omega)\hat{a}^{tr}_{\omega}+g^{ev}(\omega)\hat{a}^{ev}_{\omega}\right)\hat{\sigma}_z(t).\nonumber\\
\label{eq:eqmotion_sigma_std}
\end{eqnarray}
Formally integrating (\ref{eq:eqmotion_a}) and substituting into (\ref{eq:eqmotion_sigma_std}) yields:
\begin{eqnarray}\label{eq:stdint}
\dot{\hat{\sigma}}&=&-i\omega_A\hat{\sigma}(t)\nonumber\\
&&-\frac{i}{\hbar}\int_{0}^{\infty}\!\!\!\!d\omega\left(g^{tr}(\omega)\hat{a}^{tr}_{\omega}(0)+g^{ev}(\omega)\hat{a}^{ev}_{\omega}(0)\right)e^{-i\omega t}\hat{\sigma}_z(t)\nonumber\\
&&+\frac{1}{\hbar^2}\left(|g^{tr}(\omega)|^2+|g^{ev}(\omega)|^2\right)\int_{-\infty}^{t}\!\!\!\!\!\!d\tau\hat{\sigma}(\tau)e^{-i\omega(t-\tau)}\hat{\sigma}_z(t).\nonumber\\
\end{eqnarray}
Note that in (\ref{eq:stdint}) there are no cross-terms because the traveling creation and annihilation operators commute with the evanescent ones. 

We can repeat the same procedure using the Green function formalism. For the equation of motion of $\hat{\sigma}$ we obtain 
\begin{eqnarray}\label{eq:eqmotion_sigma_green}
\dot{\hat{\sigma}}&=&-i\omega_A\hat{\sigma}(t)-\frac{i}{\hbar}\int_{0}^{\infty}\!\!\!\!d\omega i\sqrt{\frac{\hbar}{\pi\epsilon_0}}\frac{\omega^2}{c^2}\nonumber\\
&&\times\int\!\! d^3r^{\prime}\sqrt{\epsilon^{\prime\prime}(\vec{r}^{\prime},\omega)}\vec{d}^T\bar{\bar{G}}(\vec{r}_A,\vec{r}^{\prime},\omega)\hat{\vec{f}}_{\omega}(\vec{r}^{\prime})\hat{\sigma}_z(t). 
\end{eqnarray}
By formal integration, we express the solution of (\ref{eq:fmotion}) and substitute it into (\ref{eq:eqmotion_sigma_green}). Using the Green function property (\ref{Eq:Green_property}) for nonmagnetic materials we obtain the following expression:
\begin{eqnarray}\label{eq:greenint}
\dot{\hat{\sigma}}&=&-i\omega_A\hat{\sigma}-\frac{i}{\hbar}\hat{\vec{E}}^{(+)}_{free}(\vec{r}_A,t)\vec{d}\hat{\sigma}_{z}(t)\nonumber\\
&&+\frac{1}{\hbar^2}\int_0^{\infty}\!\!\!\!d\omega\frac{\hbar\omega^2}{\pi\epsilon_0c^2}\vec{d}^T\mathrm{Im}\left[\bar{\bar{G}}(\vec{r}_A,\vec{r}_A,\omega)\right]\vec{d}\nonumber\\
&&\times\int_{-\infty}^{t}\!\!d\tau\hat{\sigma}(\tau)e^{-i\omega(t-\tau)}\hat{\sigma}_z(t)
\end{eqnarray}
which is equivalent to (\ref{eq:stdint}). Both the source and the observation points are in vacuum so $|\vec{k}|=k_0=\omega/c$. Furthermore, we have seen in Section \ref{sec:green_tensor_guided_modes} that we find the Green tensor by integrating an analytic, $k_z$- and $n$-dependent expression over $k_z$ and summing it up over $n$. With this we can rewrite (\ref{eq:greenint}) which assumes the form
\begin{eqnarray}\label{eq:greenint2}
\dot{\hat{\sigma}}&=&-i\omega_A\hat{\sigma}-\frac{i}{\hbar}\hat{\vec{E}}^{(+)}_{free}(\vec{r}_A,t)\vec{d}\hat{\sigma}_{z}(t)\nonumber\\
&&+\frac{1}{\hbar^2}\!\!\int_0^{\infty}\!\!\!\!\!\!\!d\omega\!\!\!\int_{-\infty}^{\infty}\!\!\!\!\!\!\!dk_z\!\sum_{n=0}^{\infty}\frac{\hbar\omega^2}{\pi\epsilon_0c^2}\vec{d}^T\!\mathrm{Im}\!\left[\bar{\bar{G}}^{(n)}(\vec{r}_A,\vec{r}_A,\omega;k_z)\right]\!\vec{d}\nonumber\\
&&\times\int_{-\infty}^{t}\!\!\!\!d\tau\hat{\sigma}(\tau)e^{-i\omega(t-\tau)}\hat{\sigma}_z(t).
\end{eqnarray}
Comparing (\ref{eq:stdint}) and (\ref{eq:greenint2}), separating the evanescent and traveling contributions, and exploiting the symmetry of $\mathrm{Im}\left[\bar{\bar{G}}^{(n)}(\vec{r}_A,\vec{r}_A,\omega;k_z)\right]$ in $k_z$ we can identify
\begin{eqnarray}\label{eq:coupling_plasmons}
|g^{ev}(\omega)|^2&&=2\int_{k_0}^{\infty}\!\!\!\!dk_z\sum_{n=0}^\infty\frac{\hbar\omega^2}{\pi\epsilon_0c^2}\nonumber \\ &&\times\vec{d}^T\mathrm{Im}\left[\bar{\bar{G}}^{(n)}(\vec{r}_A,\vec{r}_A,\omega;k_z)\right]\vec{d},
\end{eqnarray}
\begin{eqnarray}\label{eq:coupling_elm}
|g^{tr}(\omega)|^2&&=2\int_{0}^{k_0}\!\!\!\!dk_z\sum_{n=0}^\infty\frac{\hbar\omega^2}{\pi\epsilon_0c^2}\nonumber\\ 
&&\times\vec{d}^T\mathrm{Im}\left[\bar{\bar{G}}^{(n)}(\vec{r}_A,\vec{r}_A,\omega;k_z)\right]\vec{d}.
\end{eqnarray}
According to Fermi's golden rule, the square of the coupling strength to a given mode of the field is proportional to the spontaneous decay rate into that mode. Because of this and Eqs. (\ref{eq:coupling_plasmons}) and (\ref{eq:coupling_elm}), we can define the decay rate into the plasmon mode associated with the $n$-th cylindrical order:
\begin{eqnarray}
 \Gamma_{pl}(\omega)&=&\sum_{n=0}^{\infty}2\int_{k_0}^{\infty}\!\!\!\!dk_z\frac{2\omega^2}{\hbar\epsilon_0c^2}\vec{d}^T\mathrm{Im}\left[\bar{\bar{G}}^{(n)}(\vec{r}_A,\vec{r}_A,\omega;k_z)\right]\vec{d}\nonumber \\
&=&\sum_{n=0}^{\infty}\Gamma_{pl}^{(n)}(\omega).
\end{eqnarray}
The above decomposition into traveling and plasmonic components is strictly valid only in the absence of losses. In case of a lossy wire, the plasmon resonances will not be singularities anymore but finite peaks with a finite width. Moreover, they will no longer be the only contribution in the evanescent region: apart from the resonance peaks, we observe a broad, low-amplitude background that is due to dissipative local circulating currents. As shown in Ref. \cite{Fleischhauer-PRA-1999}, $\epsilon^{\prime\prime}\ne0$ results in non-radiative losses which, in turn, have a contribution to $|g^{tr}(\omega)|^2$, and, because of the circulating currents, also to $|g^{ev}(\omega)|^2$.
Our main interest lies in a thin, single-mode wire where we do not have higher-order plasmon modes (see Fig.\ref{fig:allowedmodes}). This allows us to resolve the plasmon mode from the circulating surface currents to a good approximation. The evanescent part of the $n>0$ cylindrical orders contains only the circulating current background. As for the $n=0$ %04022010
contribution, one can verify that there is no peak-sitting-on-a-background behaviour in $\mathrm{Im}[\bar{\bar{G}}^{(n=0)}(\vec{r}_A,\vec{r}_A,\omega;k_z)]$, but instead we have a single peak, corresponding to the plasmon mode, with an almost perfect Lorentzian shape. So, in $n=0$ order we have no, or negligible circulating surface current contribution. Thus, in the single-mode case the decay rate into the plasmons can be calculated as
\begin{equation}
 \Gamma_{pl}(\omega)=2\int_{k_0}^{\infty}\!\!\!\!dk_z\frac{2\omega^2}{\hbar\epsilon_0c^2}\vec{d}^T\mathrm{Im}\left[\bar{\bar{G}}^{(n=0)}(\vec{r}_A,\vec{r}_A,\omega;k_z)\right]\vec{d}.
\end{equation}
%
%%%%%%%%%%%%%%%%%%%%%%%%%%%%%%%%%%%%%%%%%%%%%%%%%%%%%%%%%%%%%%%%%%%%%%%%%%%%%%%%%%%%%%
\begin{figure}
 \includegraphics[width=8cm,angle=0]{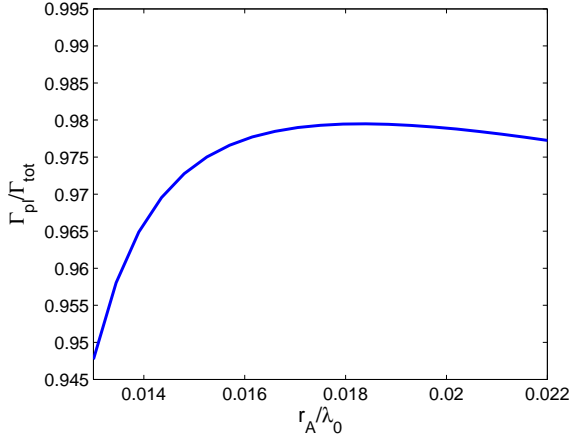}
\caption{\label{fig:plasmondecay_vs_total} Spontaneous emission rate into the plasmons relative to the total spontaneous decay rate, as a function of source distance from the wire axis $r_A$, scaled by the  vacuum radiation wavelength $\lambda_0$, in case of a wire of radius $R=0.01\lambda_0$ and $\epsilon=-75+0.6i$. There is an optimal $r_A$ at which the coupling of a single emitter to the plasmons is maximal.}
\end{figure}
%%%%%%%%%%%%%%%%%%%%%%%%%%%%%%%%%%%%%%%%%%%%%%%%%%%%%%%%%%%%%%%%%%%%%%%%%%%%%%%%%%%%%%%

Plotting the emission rate into the plasmons normalized to the overall decay rate as a function of the atomic distance from the wire axis, as seen in Fig.~\ref{fig:plasmondecay_vs_total}, we observe that there is an optimum distance at which this value is maximal. This is in accordance with the result in \cite{Chang-PRL-2006}. Our model enables us to determine the origin of this effect. If the atom gets far from the wire, intuitively, the spontaneous emission rate into the plasmons decreases rapidly, while the emission rate into the traveling modes stays more or less constant (practically zero reflection from the wire and constant vacuum spontaneous emission rate) which results in the overall decrease of $\Gamma_{pl}/\Gamma_{tot}$. However, if the atom gets closer and closer to the wire, the %04022010 changed here - I only think it is the the direct loss which really increases and become important
%traveling-wave and 
circulating current contribution increases faster than the decay rate into the $n=0$ plasmon mode. This, again, leads to the overall decrease of the relative decay rate.

%%%%%%%%%%%%%%%%%%%%%%%%%%%%%%%%%%%%%%%%%%%%%%%%%%%%%%%%%%%%%%%%%%%%%%%%%%%
%04022010 Deleted "two" here - we also discuss more
\section {Wire-mediated interaction of emitters}\label{sec:two_emitters}
%%%%%%%%%%%%%%%%%%%%%%%%%%%%%%%%%%%%%%%%%%%%%%%%%%%%%%%%%%%%%%%%%%%%%%%%%%%%

%%%%%%%%%%%%%%%%%%%%%%%%%%%%%%%%%%%%%%%%%%%%%%%%%%%%%%%%%%%%%%%%%%%%%%%
\subsection{sub- and super-radiant coupling mediated by the nano-wire}
%%%%%%%%%%%%%%%%%%%%%%%%%%%%%%%%%%%%%%%%%%%%%%%%%%%%%%%%%%%%%%%%%%%%%%%

If we place two emitters along the single-mode wire, both at the same distance from the surface, close enough to couple to the single surface plasmon mode, they will have a long-range interaction mediated by the plasmons. In the following, we will investigate %take a look at the nature of 
this interaction.
The rotating-wave Hamiltonian of two identical, two-level atoms interacting with the environment in dipole approximation assumes the form
\begin{equation}
\begin{aligned}
&\hat{H}=\int d^3\vec{r}\int_{0}^{\infty} d\omega\hbar\omega\hat{\vec{f^{\dagger}_{\omega}}}(\vec{r})\hat{\vec{f}}_{\omega}(\vec{r})+\frac{1}{2}\hbar\omega_{A}(\hat{\sigma}_{z1}+\hat{\sigma}_{z2})\\
&+\left(-\hat{\sigma}^\dagger_1\hat{\vec{E}}^{(+)}(\vec{r}_{1},t)\vec{d}_1-\hat{\sigma}^\dagger_2\hat{\vec{E}}^{(+)}(\vec{r}_{2},t)\vec{d}_2+H.c.\right)
\end{aligned}
\end{equation}
where the indices $1$ and $2$ refer to the 1st and 2nd atom.
Similarly to the treatment in \cite{Kaestel-PRA-2005}, we can write up the master equation, trace out the reservoir and after applying Born and Markov approximations we get the equation of motion for the reduced, two-atom density matrix:
\begin{equation}\label{eq:master_2atom}
\begin{aligned}
 &\dot{\hat{\rho}}=-\sum_{k,l=1}^{2}\frac{\Gamma_{kl}}{2}\left(\hat{\sigma}^\dagger_l\hat{\sigma}_k\hat{\rho}+\hat{\rho}\hat{\sigma}^\dagger_l\hat{\sigma}_k-2\hat{\sigma}_k\hat{\rho}\hat{\sigma}^\dagger_l\right)\\
&+i\sum_{k,l=1}^{2}\delta\omega_{k,l}\left[\hat{\sigma}^\dagger_l\hat{\sigma}_k,\rho\right].
\end{aligned}
\end{equation}
$\delta\omega_{kk}$ are the single-atom Lamb shifts and $\delta\omega_{12}$ is the radiative dipole-dipole shift. Similarly, $\Gamma_{kk}$ is the single-atom decay rate $\Gamma_{tot}$ seen in Section \ref{sec:single_atom} and $\Gamma_{12}$ is a contribution which describes the effect of atoms upon each other.
\begin{eqnarray}
\Gamma_{12}&=&\frac{2\omega_A^2 d_{1i} d_{2j}}{\hbar\epsilon_0c^2}\mathrm{Im}\left[G_{ij}(\vec{r}_1,\vec{r}_2,\omega_A)\right] \\
\delta\omega_{12}&=&\frac{d_{1i}d_{2j}}{\hbar\epsilon_0\pi}\mathcal{P}\int_{0}^{\infty}\!\!\!\!d\omega\frac{\omega^2}{c^2}\frac{\mathrm{Im}\left[G_{ij}(\vec{r}_1,\vec{r}_2,\omega)\right]}{\omega-\omega_A}
\end{eqnarray}
The one-atom flip operators have the form $\hat{\sigma}_k=|g\rangle_{kk}\langle e|$, with $|g\rangle_k$ being the ground state and $|e\rangle_k$ the excited state of the $k$-th atom. Since $\Gamma_{21}=\Gamma_{12}$ Eq. (\ref{eq:master_2atom}) is diagonal in the basis of symmetrized and antisymmetrized states, namely, $|ee\rangle$, $|S\rangle=(|ge\rangle+|eg\rangle)/\sqrt{2}$, $|AS\rangle=(|ge\rangle-|eg\rangle)/\sqrt{2}$ and $|gg\rangle$. Writing up the equations of motion for the populations of these states we get

\begin{eqnarray}
 \dot{\rho}_{ee}^{ee}&=&-2\Gamma_{11}\rho_{ee}^{ee} \\
\dot{\rho}_{S}^{S}&=&(\Gamma_{11}+\Gamma_{12})\rho_{ee}^{ee}-(\Gamma_{11}+\Gamma_{12})\rho_{S}^{S} \\
\dot{\rho}_{AS}^{AS}&=&(\Gamma_{11}-\Gamma_{12})\rho_{ee}^{ee}-(\Gamma_{11}-\Gamma_{12})\rho_{AS}^{AS} \\
\dot{\rho}_{gg}^{gg}&=&(\Gamma_{11}+\Gamma_{12})\rho_{S}^{S}+(\Gamma_{11}-\Gamma_{12})\rho_{AS}^{AS}.
\end{eqnarray}
Additionally, from the equations of the coherences, we perceive that the resonance frequency of $|AS\rangle$ has been shifted downwards by $\delta\omega_{12}$, whereas that of $|S\rangle$ has increased by the same amount. We used the following notation for the matrix elements of the two-particle density operator: $\rho^{gs}_{es}=\langle gs|\hat\rho|es\rangle$.
%04022010 remove "It is obvious that" saying obvious is not nice to readers who don't see it. 
The magnitude of $\Gamma_{12}$ will be responsible for the super- and subradiance of the symmetric and antisymmetric transitions. 

%%%%%%%%%%%%%%%%%%%%%%%%%%%%%%%%%%%%%%%%%%%%%%%%%%%%%%%%%%%%%%%%%%%%%%%
\begin{figure}
 \includegraphics[width=8cm,angle=0]{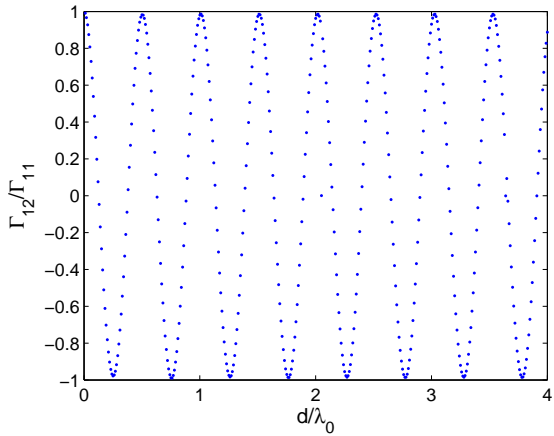}
\caption{\label{fig:superrad1} Variation of $\Gamma_{12}/\Gamma_{11}$ as a function of the distance $d$ between two emitters, in a lossless, single-mode wire of radius $R=0.01\lambda_0$. The emitter - wire axis distance is $r_A=0.015\lambda_0$. The oscillations result in the alternation between super-and subradiance of the symmetric and antisymmetric atomic transitions. All distances are scaled by the vacuum radiation wavelength $\lambda_0$. }
\end{figure}
%%%%%%%%%%%%%%%%%%%%%%%%%%%%%%%%%%%%%%%%%%%%%%%%%%%%%%%%%%%%%%%%%%%%%%
\begin{figure}
 \includegraphics[width=8cm,angle=0]{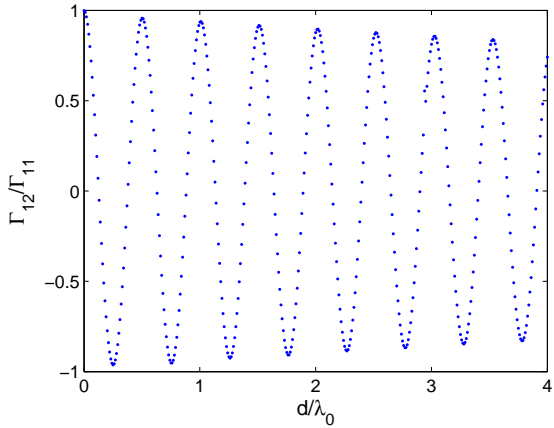}
\caption{\label{fig:superrad2} Variation of $\Gamma_{12}/\Gamma_{11}$ as a function of the distance $d$ between two emitters, in a single-mode, lossy wire ($\epsilon=-75+0.6i$) of radius $R=0.01\lambda_0$. The emitter - wire axis distance is $r_A=0.015\lambda_0$. The oscillation is damped because of the dissipation of plasmons upon propagation. All distances are scaled by the vacuum radiation wavelength $\lambda_0$.}
\end{figure}
%%%%%%%%%%%%%%%%%%%%%%%%%%%%%%%%%%%%%%%%%%%%%%%%%%%%%%%%%%%%%%%%%%%%%%%

Fig.~\ref{fig:superrad1} and Fig.~\ref{fig:superrad2} show the value of the cross relaxation rate $\Gamma_{12}$ normalized to the single-atom decay rate $\Gamma_{11}=\Gamma_{22}$, where we %04022010
have varied the distance between the two atoms by several vacuum wavelengths while keeping the atom - wire distance constant, with metal losses switched off and on, respectively. We observe an oscillatory behaviour of $\Gamma_{12}$ between $\pm\Gamma_{11}$, where the period of oscillations matches exactly the longitudinal wavelength of the surface plasmon mode. In essence,  this effect is caused by the interference of the radiation emitted by the two atoms. Depending on the distance, the two plasmon contributions emitted by the two atoms while being in the symmetric (or antisymmetric) state may either interfere constructively or destructively, giving rise to enhanced or suppressed decay rates, respectively. This means that varying the inter-atomic distance, we can induce superradiance for the symmetric transitions while suppressing spontaneous decay for the antisymmetric ones, or vice versa. Besides, %04022010 onE 
 one can see in Fig. \ref{fig:superrad1} that the maximum value of $\Gamma_{12}$ is very close to $\Gamma_{11}$ which means an efficient enhancement/suppression of sponateous decay of the corresponding transitions. Therefore, the atoms are indeed strongly coupled to the guided mode - not only is the overall decay rate greatly enhanced but almost the whole of the radiated energy is carried by the plasmons.   %04022010 
If there are losses in the metal, however, the oscillations are damped. 

\begin{figure}
 \includegraphics[width=8cm,angle=0]{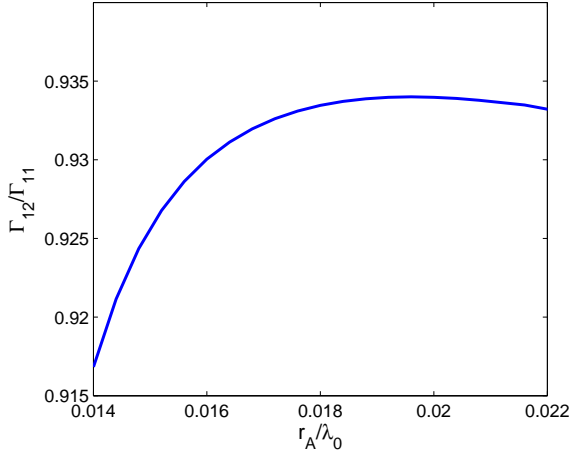}
\caption{\label{fig:optimal_pos} Magnitude of an extremal value of $\Gamma_{12}/\Gamma_{11}$ as the function of the atom - wire axis distance $r_A$, scaled by the vacuum radiation wavelength $\lambda_0$, at wire radius $R=0.01\lambda_0$ and inter-atomic distance $d=1.01\lambda_0$. We use $\epsilon=-75+0.6i$. There is an optimal $r_A$ where the coupling to the guided mode is maximal, resulting in a minimal damping of the oscillations of $\Gamma_{12}/\Gamma_{11}$.}
\end{figure}

Finite material losses, i.e. $\epsilon^{\prime\prime} \ne 0$ give rise to another effect  related to the discussion in Section \ref{subsec:sub-wavelength wire}. If we put an emitter too close to the wire surface, non-radiative losses become increasingly important.
Thus although the coupling strength to the plasmons increases, the non-radiative channel will take over.
Therefore, as Fig.~\ref{fig:optimal_pos} shows, there is, again, an optimal distance of the emitters, when the contrast of the super- and subradiance is maximal. However, the optimum distance is slightly different from the one found in Section \ref{subsec:sub-wavelength wire}. %This can be understood as follows.
Since we have two, strongly interacting emitters, the decay rate of the two-atom system into the plasmon mode will change differently as compared to the single-atom decay rate. Because of this, one can observe that the maximum in Fig.~\ref{fig:optimal_pos} is slightly shifted compared to that of Fig.~\ref{fig:plasmondecay_vs_total}. 

\begin{figure}
 \includegraphics[width=8cm,angle=0]{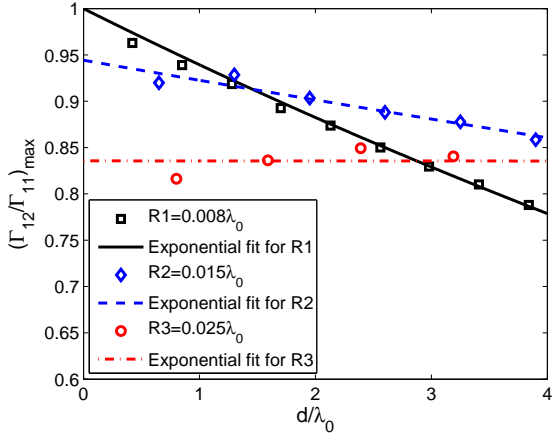}
\caption{\label{fig:optimal_R} Maximal values of $\Gamma_{12}/\Gamma_{11}$, as a function of propagation distance $d$, %04022010
for three different single-mode wire radii. We use $\epsilon^{\prime\prime}=-75+0.6i$. The lines are exponential fits to the $\Gamma_{12}/\Gamma_{11}$ peak values, represented by %04022010
the markers. As $R$ decreases, the propagation losses increase but the largest achievable  $(\Gamma_{12}/\Gamma_{11})_{max}$ value %04022010
also gets higher. Thus, at a given inter-emitter distance there exists a single $R$ that ensures a maximal strength of interaction between the emitters along the wire.}
\end{figure}

The efficiency of the coupling between two emitters separated by a fixed distance $d$ will depend on several factors. In order to maximize
the single-atom plasmon coupling the atom should be placed as close as possible to the wire surface. Since at some point non-radiative losses of
the single atom increase more rapidly when approaching the surface there is an optimum atom-wire distance for a given wire radius $R$.
On the other hand the propagating plasmons also experience absorption originating from the same material losses. These absorption losses
depend on the structure of the mode function and thus also on the wire radius $R$. 
We%04022010 
have examined whether for a fixed distance $d$ along the wire 
there is a wire radius allowing an optimal compromise between these two effects.  In Fig.~\ref{fig:optimal_R} we %04022010
plot the enveloping curve of the maxima of $\Gamma_{12}/\Gamma_{11}$ as a function of inter-atomic distance, in case of three different wire radii. In each case we used a wire-emitter distance that allowed an optimal atom-plasmon coupling. One can see that at a smaller radius $R$, the achievable maximum of $\Gamma_{12}/\Gamma_{11}$  is higher. However, as the propagation distance increases, the value of the envelope decreases 
faster compared to the case of a larger $R$. According to this, if the emitters are close to each other, it is better to have a wire with a smaller radius. On the other hand, if they are %04022010
further apart the optimal radius would be a larger one. There is thus an optimal wire radius for a given inter-atomic distance, but there is no single optimal $R$ for all cases.\\

%%%%%%%%%%%%%%%%%%%%%%%%%%%%%%%%%%%%%%%%%%%%%%%%%%%%%%%%%%%%%%%%%%%%%%%%%%%%%%%%%%%%%
%%%%%%%%%%%%%%%%%%%%%%%%%%%%%%%%%%%%%%%%%%%%%%%%%%%%%%%%%%%%%%%%%%%%%%%%%%%%%%%%%%%
%%%%%%%%%%%%%%%%%%%%%%%%%%%%%%%%%%%%%%%%%%%%%%%%%%%%%%%%%%%%%%%%%%%%%%%%%%%%%%
\begin{figure}
\includegraphics[width=8cm,angle=0]{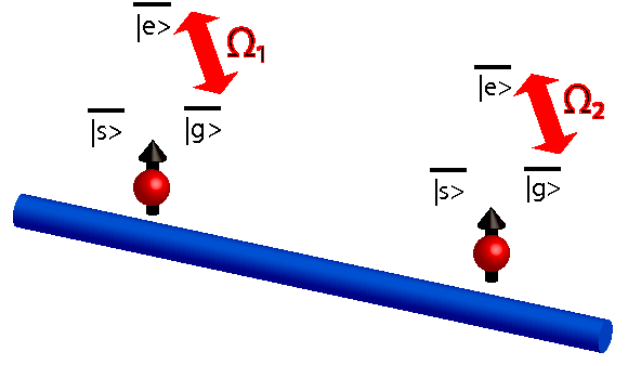}
\caption{\label{fig:phase_gate_setup} Realization of a deterministic quantum phase gate, by coupling the $|e\rangle-|g\rangle$ of two lambda atoms to a single surface plasmon mode, inducing superradiance in the two-atom system, and applying external classical $2\pi$ pulses. }
\end{figure}
%%%%%%%%%%%%%%%%%%%%%%%%%%%%%%%%%%%%%%%%%%%%%%%%%%%%%%%%%%%%%%%%%%%%%%%%%%%%%%%%%%%%%%%
\begin{figure}
 \includegraphics[width=8cm,angle=0]{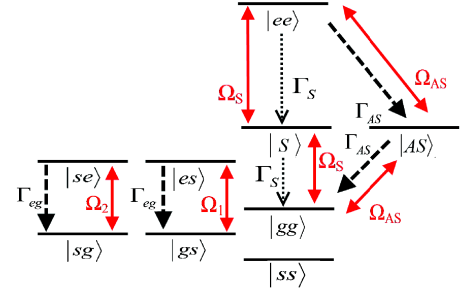}
\caption{\label{fig:phase_gate_levels} Effective level scheme of two, externally driven, interacting 3-level lambda atoms, seen in Fig.~\ref{fig:phase_gate_setup}. $\Omega_S$ and $\Omega_{AS}$ are, respectively, the symmetric and antisymmetric superpositions of $\Omega_1$ and $\Omega_2$. Because of the strong coupling to a single plasmonic mode, a high-contrast superradiance effect is present in the system, inducing a slow decay in the symmetric transitions and a fast one in the antisymmetric transition.}
\end{figure}
%%%%%%%%%%%%%%%%%%%%%%%%%%%%%%%%%%%%%%%%%%%%%%%%%%%%%%%%\subsection{plasmon-mediated quantum phase gate}
%%%%%%%%%%%%%%%%%%%%%%%%%%%%%%%%%%%%%%%%%%%%%%%%%%%%%%%%%%%%%

As a possible application, in the following we are going to analyze the realization of a deterministic quantum phase gate between two lambda atoms coupled by a single surface plasmon mode of the nanowire, as shown  in Fig~\ref{fig:phase_gate_setup}. We assume that only the $|g\rangle\rightarrow|e\rangle$ transition of each emitter is coupled to the single guided mode of the wire. If the two emitters are several wavelengths apart they can individually be addressed by lasers coupling the $|g\rangle\rightarrow|e\rangle$ transition. The phase gate works by exploiting the large difference in the decay rates between the super- and sub-radian states shown in Fig. \ref{fig:superrad2}. Assume first that we can ignore the decay of the excited level $|e\rangle$ in the $|g\rangle-|e\rangle$ two level system. In this case it is well known  that a $2\pi$ pulse, forcing the system to do a full Rabi oscillation, gives an additional $\pi$ phase to the atomic system. On the other hand if the decay rate of the excited state $\Gamma_{eg}$ is much stronger than the resonant Rabi frequency $\Omega\ll\Gamma$, the driving field cannot induce Rabi oscillations between the two levels, and the drive merely introduces a scattering rate $\propto\Omega^2/\Gamma_{eg}$, which vanishes in the limit of weak driving or strong decay. The idea behind the phase gate is that due to the subradiant effect we saw above, the interaction of a nearby emitter can change the excited state  from decaying into non-decaying thereby allowing the first emitter to pick up a $\pi$ phase shift conditioned on the other other .

Specifically, consider the level scheme of the two-atom system shown in Fig~\ref{fig:phase_gate_levels}, where we use the symmetric-antisymmetric superposition for the transitions coupled to the wire. The  external, classical fields with Rabi frequencies $\Omega_1$ and $\Omega_2$ give rise to coupling strengths to the symmetric or antisymmetric states given by
\begin{eqnarray}
 \Omega_S=\frac{1}{\sqrt{2}}(\Omega_1+\Omega_2)\\
\Omega_{AS}=\frac{1}{\sqrt{2}}(\Omega_1-\Omega_2)
\end{eqnarray}
Assume now that the distance between the atoms is such that the symmetric state $|S\rangle$  is subradiant, whereas the  antisymmetric state $|AS\rangle$ is superradiant $\Gamma_{AS}\gg\Gamma_S$ and that $\Omega_2=\Omega_1$ corresponding to $\Omega_{AS}=0$ and $\Omega_{S}=\sqrt{2}\Omega_1$. Choosing the driving strengths $\Omega_S$ to be in between the two decay rates 
\begin{equation}
\Gamma_{AS}\gg \Omega_S \gg\Gamma_S
\label{eq:inbetween}
\end{equation}
ensures that transition to the double excited state $|ee\rangle$ is blocked by the strong decay $\Gamma_{AS}$, whereas we can perform a $2\pi$ Rabi oscillation on the two level system given by $|gg\rangle$ and $|S\rangle$ (see Fig. \ref{fig:phase_gate_levels}). This thus ensures that we achieve a phase change of $\pi$ on the $|gg\rangle$ state. Starting from the states $|gs\rangle$ and $|sg\rangle$ there is essentially only a single atom interaction with the wire. The excited state therefore always have a fast decay such that the driving is too weak to excite the atoms. Furthermore the state $|ss\rangle$ is completely unaffected by the classical light pulses and in the ideal 
limit the entire process will therefore have the truth table 
\begin{eqnarray}
|ss\rangle&\rightarrow &|ss\rangle\nonumber \\
|sg\rangle&\rightarrow &|sg\rangle\nonumber \\
|gs\rangle&\rightarrow &|gs\rangle\nonumber \\
|gg\rangle&\rightarrow &-|gg\rangle.
\end{eqnarray}

In reality the finite ratio $\Gamma_{AS}/\Gamma_{S}$ will limit how well one can fulfill the condition in Eq. (\ref{eq:inbetween}) and this will limit the fidelity $F$ of the gate. Choosing the Rabi frequency too small will lead a decay from the excited state during the $2\pi$ pulse on the $|gg\rangle-|S\rangle$ transition. This will result in an error with a probability $\propto\Gamma_S T_{{\rm pulse}}\propto\Gamma_S/\Omega_1$, where the pulse duration $T_{{\rm pulse}}$ is set  by $T_{{\rm pulse}}\sim 1/\Omega_1$. Choosing the driving too strong will allow scattering from the transitions which are supposed to be blocked by the fast decay of the excited states with decay rates $\Gamma_S\propto \Gamma_{eg}$. This introduces an error  probability $\propto T_{{\rm pulse}}\Omega_1^2/\Gamma_{eg} \propto \Omega_1/\Gamma_{eg}$ resulting in a total imperfection
\begin{equation}\label{eq:imperfection}
1-F\sim \frac{\Omega_1}{\Gamma_{eg}}+\frac{\Gamma_{S}}{\Omega_{1}}.
\end{equation}
Minimizing (\ref{eq:imperfection}) with respect to $\Omega_1$, we find that the minimal imperfection scales as
\begin{equation}\label{eq:imperfection_min}
 1-F_{opt}\sim \sqrt{\frac{\Gamma_{S}}{\Gamma_{eg}}}.
\end{equation}

\begin{figure}
\includegraphics[width=8cm,angle=0]{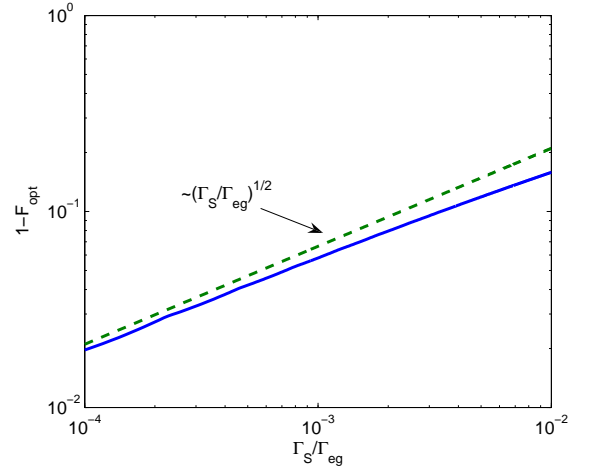}
\caption{\label{fig:fid_error_loglog} Fidelity of a maximally entangled state created by the phase gate optimized with respect to the drive strength. The phase gate fidelity error scales roughly as $\sqrt{\Gamma_S/\Gamma_{eg}}$  (dashed line) in the limit $F\rightarrow 1$.}

\end{figure}

To confirm this rough scaling analysis we show in 
Fig~\ref{fig:fid_error_loglog} the result of a full numerical simulation of the density matrix equation for the system. In the figure we show the fidelity $F=\langle \psi_{{\rm ideal}}|\rho  |\psi_{{\rm ideal}}\rangle$ optimized over the driving strength $\Omega_1$ as a function of the ratio of the decay rates $\Gamma_S/\Gamma_{eg}$. Here $|\psi_{{\rm ideal}}\rangle$ is a maximally entangled state created using the phase gate combined with ideal $\pi/2$ pulses  on the $|g\rangle-|s\rangle$ transitions. 
The dashed line shows the scaling derived  above $1-F=c \sqrt{ \Gamma_{S}/\Gamma_{eg}}$, with the constant $c$ chosen to match the behavior as $F\rightarrow 1$. 

\begin{figure}
 \includegraphics[width=8cm,angle=0]{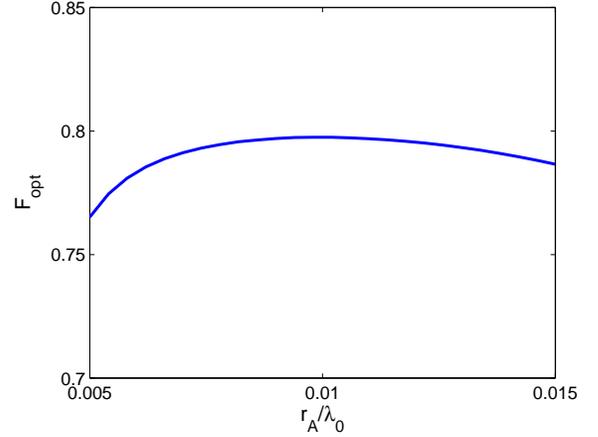}
\caption{\label{fig:fidelity} %04022010 changed caption and moved the figure
Optimal fidelity of a maximally entangled state generated with the phase gate  as a function of the atom - wire axis distance $r_A$. $r_A$ is scaled with the vacuum radiation wavelength $\lambda_0$. The two-atom states $|ss\rangle$, $|sg\rangle$, $|gs\rangle$ and $|gg\rangle$ are initially equally populated. We used parameters $R=0.003\lambda_0$, $d=0.08\lambda_0$, $\epsilon=-75+0.6i$ in the simulation.}
\end{figure}

The analysis above  has not been been tied to any particular system, but applies to to any system where there is a large difference between the super- and sub-radiant states.  We now turn to the specific implementation in terms of atoms coupled through the plasmon. As we saw above, the subradiant state do not have a completely vanishing decay rate  and accordingly this will limit the fidelity of the gate.
In Fig.~\ref{fig:fidelity}, we show the optimal fidelity %($\mathrm{Tr}[\hat{\rho}_{ideal}\hat{\rho}_{final}]$) 
of the gate operation, optimized by varying $\Omega_S$,  as a function of the atom - wire axis distance $r_A$. For the simulation, we used $R=0.003\lambda_0$ and $d=0.08\lambda_0$. 

%%%%%%%%%%%%%%%%%%%%%%%%%%%%%%%

\begin{figure}
\includegraphics[width=8cm,angle=0]{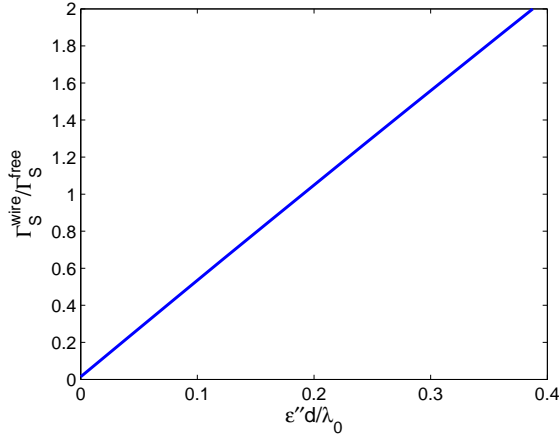}
\caption{\label{fig:sym_wire_free} Ratio of wire-loss-induced and free-space-loss-induced contributions of $\Gamma_S$, as a function of $\epsilon^{\prime\prime}d/\lambda_0$ where we restrict $d$ to take up values of the distances belonging to the minima of $\Gamma_{12}/\Gamma_{11}$ (see Fig.\ref{fig:superrad2}). In the region of $\Gamma_S^{wire}/\Gamma_S^{free}<1$ the gate fidelity could be considerably increased by suppressing the spontaneous radiation into free space.}
\end{figure}

The results show a maximum fidelity of 80\% for the parameters above. There are two major detrimental effects which reduce the superradiance-subradiance contrast and, consequently, the gate fidelity, namely, the coupling to free space and the wire losses. Due to the free space coupling, a small amount of the radiated energy will be scattered into the far-field region, resulting in a non-zero contribution to $\Gamma_{S}$. The wire losses induce traveling-wave and circulating current dissipation - leading to an additional, local loss of energy. 
% I don't like this discussion about width. I think that it is much more intuitive to just discuss loss.
%- and a broadening of the plasmonic resonance. Because of the broadening, the plasmonic field contains a range of $k_z$ values around $k_z^{pl}$. However, the condition for subradiance is strictly fulfilled only for $k_z=k_z^{pl}$. Thus, for some parts of the plasmonic radiation, there will be a phase mismatch, resulting in an imperfect subradiance, i.e. some energy will leak out of the inter-emitter wire section. 
This also  %effect, along with the local non-radiative losses,
 contributes  to $\Gamma_{S}$. 
It is possible to suppress the spontaneous emission into free space, for example, by placing the atoms into a photonic bandgap material. This would considerably enhance the gate fidelity only if the free-space contribution to $\Gamma_{S}$ is dominant with respect to the wire-loss contribution. Fig.~\ref{fig:sym_wire_free} shows the ratio $\Gamma_{S}^{wire}/\Gamma_S^{free}$ as a function of $d\epsilon^{\prime\prime}/\lambda_0$, for distances $d$ corresponding to the minima of $\Gamma_{12}/\Gamma_{11}$. As one can see, to increase the inter-emitter distance $d$ while keeping $\Gamma_{S}^{wire}/\Gamma_S^{free}<1$, we need to decrease $\epsilon^{\prime\prime}$. For the given parameters, this means that, for instance, for $d\approx2\lambda_0$ we have to have $\epsilon^{\prime\prime}\approx 0.1$. 
Another means of increasing the fidelity could be to build a cavity for the plasmon, e.g., by using a ring shaped wire, but this will depend on the losses of such a cavity mode, e.g., due to radiation. A full investigation of this is beyond the scope of this paper.   %avoid %the energy loss due to the phase mismatch of the $k_z$ components, by using a metallic ring instead of an infinite wire. Thus, one would essentially get a pair of atoms strongly coupled to a ring resonator. \\

%%%%%%%%%%%%%%%%%%%%%%%%%%%%%%%%%%%%%%%%%%%%%%%%%%%%%%%%%%%%%%

%\subsection{Geometry-free analysis of the phase gate}\label{subsec:geomfree_analysis}

%We would like to briefly describe the asymptotic behaviour and scaling of the Rabi-frequency-optimized fidelity $F_{opt}$ of the phase gate mentioned above, as $F_{opt}\rightarrow 1$, i.e. as $\Gamma_S/\Gamma_{eg}\rightarrow 0$. This description is independent of the particular physical system used to realize the gate. 

%The gate imperfection $1-F$ is proportional to the unwanted population scattering from certain states of the system. There are essentially two mechanisms leading to imperfection. One of them is the scattering of population from the fast decaying states $|se\rangle$ and $|es\rangle$. The population transferred to these states is proportional to $\Omega_1^2/\Gamma_{eg}^2$. The population scattered during interaction goes like $(\Omega_1^2/\Gamma_{eg}^2)\Gamma_{eg}T_{pulse}$ where $T_{pulse}\propto 1/\Omega_1$ is the pulse duration. So the imperfection is proportional to $\Omega_1/\Gamma_{eg}$.
%The other source of imperfection is the population scattered from the state which does the $2\pi$ cycle, i.e. $|S\rangle$. With a similar logic as above, we see that the error resulting from this mechanism is proportional to $\Gamma_{S}/\Omega_{S}$. This, if $\Gamma_S/\Gamma_{eg}\rightarrow 0$, goes like $\Gamma_{S}/\Omega_1$. So the total imperfection scales as
%

%%%%%%%%%%%%%%%%%%%%%%%%%%%%%%%%%%%%%%%%%%%%%%%%%%%%%%%%%%%%%%%%%%%%%%%%%%%%%%%%%%%%%%%
\subsection{realization of spin boson model}
%%%%%%%%%%%%%%%%%%%%%%%%%%%%%%%%%%%%%%%%%%%%%%%%%%%%%%%%%%%%%%%%%%%%%%%%%%%%%%%%%%%%%%%

\begin{figure}
\includegraphics[width=6.5cm,angle=0]{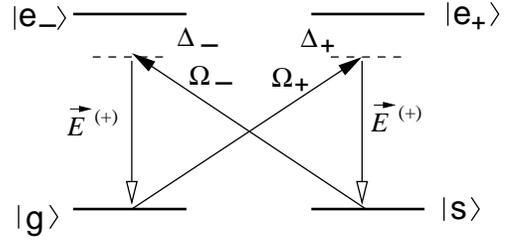}
\caption{\label{fig:spin-boson} Raman coupling scheme for the realization of a spin boson model.
In the rotating wave approximation%04022010 deleted the following which is rather obvious " well justified for single-photon optical transition,"
 the
atom can make a transition from $|g\rangle$ to $|s\rangle$ either by emitting a photon into
$\hat E$ via the transition $|g\rangle \to |e_+\rangle \to |s\rangle$ or by absorbing a photon
from $\hat E$ via the transition $|g\rangle \to |e_-\rangle \to |s\rangle$.
}
\end{figure}

The plasmon mediated interaction between spins along a nano-wire could also be used
to implement interesting and important many-body models such as the spin boson
model \cite{Zwerger-review}, in which a chain of spins is coupled to a one-dimensional 
continuum of bosonic
modes through an interaction of the type
\begin{equation}
\hat H_{\rm SB} = \int_0^\infty\! \!{\rm d}\omega \, \sum_j \left(\hat \sigma_j 
+\hat\sigma_j^\dagger\right)\Bigl(g(\omega) \hat a_\omega + g^*(\omega) \hat a_\omega^\dagger\Bigr).
\end{equation}
Different from the atom-plasmon coupling (\ref{Eq:dipole_interaction}) discussed throughout this
paper, this Hamiltonian contains counter rotating terms of the form $\hat\sigma^\dagger \hat a^\dagger$ and
$\hat\sigma \hat a$. Typically these terms can be neglected for electromagnetic modes oscillating
at optical frequencies. However it is possible to compensate the fast oscillations by using
effective two-photon Raman transitions rather than single-photon ones \cite{Carmichael}. 
Consider e.g. the $F=1/2$--$F=1/2$ coupling scheme shown in Fig.\ref{fig:spin-boson}, where the lower
state $|g\rangle$ is coupled to another state $|s\rangle$ in the ground-state manifold via a two-photon Raman transition through an excited state $|e_+\rangle$ involving a right cicular polarized external drive field  of Rabi-frequency $\Omega_+$ and the linear polarized quantized plasmon field ${\vec E}$. In addition state $|s\rangle$ is coupled through
another excited state $|e_-\rangle$ via a two-photon Raman transition involving a left circular polarized external drive field
$\Omega_-$ and again the linear polarized plasmon field ${\vec E}$. 
It is assumed that the Raman transitions are in two-photon resonance. On the other hand
the single-photon transitions shall have a large detuning $\Delta_\pm$ such that the excited states
can be adiabatically eliminated.
For each single-photon transition the rotating-wave approximation is well justified since we are
considering optical fields. 
\begin{eqnarray}
\hat H_I &=& \sum_j \Bigl\{\hbar \Omega_+^* \hat\sigma_{e+,g}^j 
+ \Omega_-^* \hat\sigma_{e-,s}^j +h.a.\Bigr\} \\
& +&\sum_j \left(\hat \sigma_{e-,g} \vec d_-\cdot\hat{\vec E}^{(+)}(z_j) +\hat \sigma_{e+,s} \vec d_+\cdot\hat{\vec E}^{(+)}(z_j)
\right).\nonumber
\end{eqnarray}
Here $\hat \sigma_{\mu,\nu}^j=|\mu\rangle_{jj}\langle\nu|$ are the atomic flip operators between
states $|\nu\rangle$ and $|\mu\rangle$ of the $j$th atom at position $z_j$, and $\vec d_\pm$ are the vector dipole 
moments of the transitions $|s\rangle -- |e_+\rangle$ and $|g\rangle -- |e_-\rangle$ respectively.
Adiabatically eliminating the excited states then leads besides some ac-Stark shifts to the 
spin-boson interaction
\begin{equation}
\hat H_{\rm eff} \sim \sum_j\Bigl(\frac{\Omega_+^* {\vec d}_+}{\Delta_+} \sigma_{s,g}^j \,
\hat{\vec E}^{(-)}(z_j)
+\frac{\Omega_-^* {\vec d}_-}{\Delta_-} \sigma_{g,s}^j \hat{\vec E}^{(-)}(z_j)+ h.a\Bigr).
\end{equation}

%%%%%%%%%%%%%%%%%%%%%%%%%%%%%%%%%%%%%%%%%%%%%%%%%%%%%%%%%%%%%%%%%%%%%%%%%%%%%%%%%%%%%%%
\section{Summary}\label{sec:summary}
%%%%%%%%%%%%%%%%%%%%%%%%%%%%%%%%%%%%%%%%%%%%%%%%%%%%%%%%%%%%%%%%%%%%%%%%%%%%%%%%%%%%%%%

In the present paper we have analyzed the coupling of quantum dipole oscillators to the plasmon modes
of an infinite cylindrical nano-wire by means of a Green function approach taking into account wire losses. 
Placing a single emitter close to the surface of a cylindrical metallic wire with sub-wavelength radius, there is a strong interaction between the emitter and the traveling surface plasmon modes of the metal. The magnitude of the interaction becomes apparent in a substantial increase of the total spontaneous decay rate going up to several hundred times the vacuum value. Decreasing the wire radius, only the basic cylindrical mode survives with a rapidly increasing radial confinement, allowing for a single-mode wire and an increasingly strong emitter-plasmon interaction. 
We have calculated the spectrum of the plasmon modes and analyzed their broadening in the presence of metal
losses ($\epsilon^{\prime\prime}\ne0$) for different wire radii $R$. For fixed $\epsilon^{\prime\prime}$, the broadening and so the propagation losses increase with decreasing wire radius since a growing part of the field is concentrated inside the wire. For a sub-wavelength
nano-wire the dependence of the resonance width on the wire radius turns from an $R^{-1/2}$ to an 
$R^{-3/2}$ behaviour. We derived explicit expressions for the coupling strength of an atom to the
plasmon modes and discussed its relation to non-plasmonic couplings in dependence on the wire parameter
and the atom-wire distance. For non-vanishing material losses there is an optimum emitter-wire distance for which the decay rate into the plasmons relative to the total decay rate is maximal. We also discussed the plasmon-mediated
coupling between two quantum dipoles. Putting two emitters close to the wire results in  Dicke-sub- and superradiance 
even when the separation of the emitters along the wire exceeds the vacuum wavelength by an order of magnitude. Depending on the precise distance between the emitters, the decay rate of the symmetric or the anti-symmetric Dicke state is enhanced or suppressed. Similar to the one-atom case, due to the finite material losses there is an optimum emitter-surface distance yielding the maximal coupling between the atoms. Also, for a given inter-atomic distance there exists an optimal wire radius for which the coupling between the emitters is maximal.
The possibility to control the symmetric and anti-symmetric decay rates allows to build two-qubit quantum gates using the setup. We proposed and analyzed a scheme for a deterministic two-qubit phase gate. The achievable gate fidelity for realistic parameters is on the order of 80\%. This can be attributed to two loss mechanisms: wire absorption of plasmons and spontaneous emission into free-space radiation modes. For small values of $\epsilon^{\prime\prime}$ the fidelity could be improved by embedding the atoms in a photonic band-gap material, suppressing the free-space spontaneous emission. Furthermore, introducing a wire ring instead of an infinite cylinder, the effect from the loss of excitation along the wire may be reduced while retaining the strong coupling. Additionally, a geometry-independent analysis shows that, asymptotically, the gate fidelity error scales as $\sqrt{\Gamma_S/\Gamma_{eg}}$, i.e., the second root of the ratio between the symmetric (subradiant) and one-atom decay rate, as $F\rightarrow 1$. Finally atoms coupled to plasmonic
nano-wires may also be an interesting system for the realization of interesting spin models such as the
spin-boson model.

\section {Acknowledgements}

David Dzsotjan acknowledges financial support by the EMALI Marie-Curie Network and by the Research Fund of the Hungarian Academy of Sciences (OTKA) under contract No. 78112. Anders S. S\o rensen acknowledges the support of the Villum Kann Rasmussen foundation and the Danish National Science Research Foundation. 
%If we write it like this it sounds like he should be an author.....
The authors would like to thank P. Rabl for discussions and his valuable contribution to Section III B. %\ref{subsec:geomfree_analysis}.

\end{document}